\documentclass[twocolumn,showpacs,preprintnumbers,amsmath,amssymb,
nofootinbib,superscriptaddress]{revtex4}

\usepackage{graphicx}

\def\square{\kern1pt\vbox{\hrule height 1.2pt
\hbox{\vrule width 1.2pt\hskip 3pt
\vbox{\vskip 6pt}\hskip 3pt\vrule width 0.6pt}
\hrule height 0.6pt}\kern1pt}
\def\ltwid{\mathrel{\raise.3ex\hbox{$<$\kern-.75em\lower1ex\hbox{$\sim$}}}}

\begin{document}

\title{A useful guide for gravitational wave observers to test modified gravity models}

\author{E. O. Kahya }
\email[]{emre-onur.kahya@uni-jena.de} 
\affiliation{Theoretisch-Physikalisches Institut, Friedrich-Schiller-Universit\"at Jena, 
Max-Wien-Platz 1, D-07743 Jena, Germany}
\date{\today}

\begin{abstract}

We present an extension of a previously suggested test of all modified theories of gravity that
would reproduce MOND at low accelerations. In a class of models, called ``dark matter emulators,''
gravitational waves and other particles couple to different metrics. This leads to a detectable time lag
between their detection at Earth from the same source. We calculate this time lag numerically for any event that
occurs in our galaxy up to 400 kpc, and present a graph of this possible time lag. This suggests that, 
gravitational wave observers might have to consider the possibility of extending their analysis to non-coincident
gravitational and electromagnetic signals, and the graph that we present might be a useful guideline for this effort.

\end{abstract}


\maketitle

%
%%%%%%%%%%%%%%%%%%%%%%%%%%%%%%%%%%%%%%%%%%%%%%%%%%%%%%%%%%%%%%%%%%%%%%%%%%%%%%%
%  MAIN TEXT
%%%%%%%%%%%%%%%%%%%%%%%%%%%%%%%%%%%%%%%%%%%%%%%%%%%%%%%%%%%%%%%%%%%%%%%%%%%%%%%
%

%\section{Introduction}

\section{Introduction}

 The last forty years have been a very active and an exciting period for cosmologists.
The breakthrough was the discovery of cosmic microwave background radiation (CMBR), 
an open window to the very early stages of the universe \cite{PW}. We realized that the universe is homogeneous and isotropic on large scales \cite{Mather,Smoot}. And very recently we learned that the universe is accelerating contrary to 
the common belief \cite{Riess, Perl}.

Another very exciting observational discovery was the peculiar shape of the so called 
``galaxy rotation curve''. Instead of the velocity of stars orbiting spiral galaxies
being inversely proportional to the distance away from the center of the galaxy, it was 
observed to stay almost constant \cite{RTF, RFT1, RFT2} after certain point. Then even 
more suprising observational fact was the excessive amount of distortion in the background 
galaxies near a foreground mass \cite{TVW,FKSW,SKFBW,CLKHG,Mellier,WTKAB}. Assuming that there 
is only baryonic matter in the universe, it is not possible to explain these two observations with Einstein's equations,
\begin{equation}
 {G}_{\mu \nu} =
\frac{8\pi G}{c^4} T_{\mu\nu} \; . \label{Ein} \end{equation}

Two of the following can be done to avoid this discrepancy. We can modify the right 
hand side by assuming additional dark matter, that would interact only gravitationally. Or we 
can modify the left hand side of the equation, namely the Einstein's tensor. The latter 
approach suggests to modify general relativity on cosmological scales. And it was partially
achieved in 1983 by Milgrom \cite{Milg} with the so called ``MOND'' for Modified Newtonian Dynamics.
Partially, because it is not a modification of general relativity but Newtonian dynamics on cosmological scales. 
The modification of GR would then be the underlying fundamental theory that would give 
MOND in the non-relativistic limit. 

In order to understand Eq.(\ref{Ein}), one needs to be able to handle both sides of the equation. That means 
identifying dark matter directly, by a non-gravitational experiment. Therefore direct detection is crucial 
for dark matter. For more than 30 years, experimentalists have been trying to detect dark matter without any success. 
Very recently some excitement arose about the detection, but turned out to be almost no signal with excluding new parameter space\cite{CDM}.
On the other hand, the early relativistic \cite{BM}, and the Scalar-Tensor \cite{San} formulations of MOND led to, so called 
TeVeS for ``Tensor-Vector-Scalar'', which was recently proposed by Bekenstein \cite{Bek}.
It reproduces the MOND in the non-relativistic limit and does a good job on cosmological evolution, strong gravitational lensing as well as post-Newtonian parameters \cite{Bek2,Chen}. But weak lensing is problematic in the case of colliding clusters, such as the Bullet Cluster \cite{TC,FFB,Car,AFZ}. The observed third peak of new 5-year WMAP data\cite{Dun} is also problematic in the context of MOND-like models. But still one can not claim that these models are ruled out \cite{San2,Bek3}. Thefore it is very important to find a way of testing these two explanations by means of observation.

The possibility for comparing dark matter with MOND-like models by means of a model-independent test was suggested
before\cite{KW,K,DKW}. It was proposed that one would see a time lag between arrival times of gravitational waves and other massless particles.
And the time lag was calculated for three different sources; a particular gamma-ray burst(GRB) GRB070201, 
the low mass x-ray binary(LMXB) SCo-X1, and the supernova SN1987a. Other proposed tests of TeVes with gravity waves are discussed in
\cite{Sot} and references therein. In Sec. 2 we summarize the formulation of this suggested test.
In Sec. 3 we extend this calculation numerically to any source in Milky Way galaxy. In Sec. 4 we discuss the consequences of this effect to the externally triggered searches for GWs. In Sec. 5  we investigate the ambiguities in this calculation, and see how sensitive the time lag is to the choice of using different data sets. Our conclusions are summarized in Sec. 6.
\vskip 0.3in

\section{Formulation}

TeVeS is an impressive achievement but a natural question arises when one thinks about multiple metric formalisms. 
Why do we need to have multiple metrics? It was realized that\cite{SW2} if one wants to generalize MOND, to get Tully-Fisher relation\cite{TF}
and sufficient lensing, to a stable, relativistic, covariant pure metric theory without dark matter,
it would not be possible. This no-go theorem restricts the modified gravity models,
that are consistent with Tully-Fisher relationship and the observed amount of lensing. One can violate one of the assumptions of the
no-go theorem by introducing another metric which would carry the MOND force. These class of modified gravity models with multiple metrics, which would give MOND in the non-relativistic limit, were called as ``dark matter emulators''. 
TeVeS and the scalar-vector-tensor gravity theory that was proposed by Moffat\cite{JWM,BrM} 
are examples of dark matter emulators. These models all have the following property:
All the ordinary matter should move on a geometry as if there is dark matter, 
whereas the gravity waves move on a geometry without any. 
The total time needed for a gravitational wave to reach the Earth from
a given source is the sum of distance divided by speed of light
and additional delay due to gravitatiional potential of intervening
matter along the line of sight (also known as Shapiro delay) \cite{Shap,Lon}. This can be 
restated as follows: Gravity waves couple to the usual metric of GR without
dark matter and the rest of the matter couples to a different metric 
that would be produced with GR plus dark matter. So the answer to the question at the beginning
of this section is, we might have to introduce multiple metrics for a modified gravity model
in order to explain Tully-Fisher relationship and the observed amount of lensing consistently, 
without dark matter.

  Heuristically the calculation can be understood with a simple picture.
Let us compare the arrival times of two massless particles that are produced 
by a distant source. The gravity wave is going to move on a geometry where 
there is only ordinary matter, and the other ordinary massless particle, 
which can be a photon or a neutrino, will move in a geometry where there is 
dark matter as well as the ordinary matter. Because of this additional mass
the ordinary massless particle would have to travel a geometry as if the distance 
is longer, which would mean that gravity waves arrive earlier than all the rest of 
the massless particles. In other words,, because of this additional mass, the 
ordinary massless particles feel an extra Shapiro delay as compared to gravity waves. 
Therefore the procedure is the following: Calculate the arrival time 
of a massless particle by solving the geodesic equation with a mass density being 
just that of dark matter, which would be equal to the difference of 
arrival times of these two massless particles.

 We will calculate the time lag in the context of bimetric theories mimicking 
cold dark matter which has a spherically symmetric distribution. One can express 
the time lag of an event in terms of its initial $(\vec{x}_1)$ and final $(\vec{x}_2)$ positions, by solving corresponding geodesic equations perturbatively. The result \cite{DKW} for the time lag is,
\begin{eqnarray}
\lefteqn{c \Delta t = \frac{\Delta \vec{x} \cdot \vec{x}_1}{2 \Delta x} \, 
\Delta B(r_1) - \frac{\Delta \vec{x} \cdot \vec{x}_2}{2 \Delta x} \, 
\Delta B(r_2) } \nonumber \\
& & \hspace{0.5cm} + \int_{r_2}^{r_1} \!\!\! dr \, \frac{2 G M(r)}{c^2 r}
\sqrt{1 - \frac{r_1^2 \Delta x^2 - (\vec{x}_1 \cdot
\Delta \vec{x})^2}{r^2} } \; , \qquad \label{cdt}
\end{eqnarray}
where 
\begin{equation}
\Delta B(r) = -\frac{2 G}{c^2} \, \frac{M(r)}{r} - \frac{2 G}{c^2} \!\!\!\int_r^{\infty} \!\!\!\!\! dr' \, 
\frac{M(r')}{r'} \; . \label{DBfromDA}
\end{equation}
$M(r)$ is the mass function,
\begin{equation}
M(r) \equiv 4 \pi \int_0^r dr' \, \rho(r') \; . \label{mass}
\end{equation}
Therefore, in order to mimick dark matter one has to choose a dark matter density profile $\rho(r)$ and that
will determine the amount of time lag.

\section{Calculation}

In this Section, we revisit a calculation that was done in a previous work. 
The time lag of photons from GRB070201\cite{DKW}, a short duration 
Gamma Ray Burst (GRB) which is believed to have happened at the 
Andromeda Galaxy, was calculated. To evaluate the total time lag, one needs to 
include contributions from both the Milky Way and Andromeda galaxies. In the mentioned work,  
the data for the density profiles of these two galaxies were taken from two different groups.
Therefore we decided to do the calculation again, based on the data of 
a group where they did a numerical work on the density profile of both of the galaxies \cite{KZ}.

As stated above, to calculate the time lag we need to know the form of the dark matter density profile. The NFW profile
can be used for this purpose, which assumes the following density profile of dark matter halos in numerical simulations
\cite{NFW},
\begin{equation}
\rho(r) = \frac{\rho_0}{\frac{r}{r_0} [1 + \frac{r}{r_0}]^2} \; . \label{rho}
\end{equation}

The parameters, that we need to calculate the time lag, are $\rho_0$ the characteristic density, 
and $r_0$ the radius of the halo. The summary of the Shapiro delays for GRB 070201 is given in Table \ref{Datasets}.
One can notice that the time lag becomes larger with increasing virial mass of the Andromeda galaxy.

\begin{table}[h!]

\vbox{\tabskip=0pt \offinterlineskip
\def\tablerule{\noalign{\hrule}}
\halign to245pt {\strut#& \vrule#\tabskip=1em plus2em& \hfil#\hfil&
\vrule#& \hfil#\hfil& 
\vrule#& \hfil#\hfil& \vrule#& \hfil#\hfil& \vrule#\tabskip=0pt\cr
\tablerule 
\omit&height4pt&\omit&&\omit&&\omit&&\omit&\cr
&&$\!\!\!\!{\rm Data\ Set}\!\!\!\!\!\!\!$ &&$\!\!\!\!\!{\rho (GeV/cm^3)}
\!\!\!\!\!\!$ && $\!\!\!\!\! {M_{vir}(M_{\odot})} \!\!\!\!\!$ && $\!\!\!\!\! {\Delta t(days)}
\!\!\!\!\!$ & \cr
\omit&height4pt&\omit&&\omit&&\omit&&\omit&\cr
\tablerule
\omit&height4pt&\omit&&\omit&&\omit&&\omit&\cr
&& $\!\!\!\!\! {\rm MW-Klypin\ [39] } \!\!\!\!\!\!\!$
&& $\!\!\!\!\! 0.185  \!\!\!\!\!\!\!$
&& $\!\!\!\!\! 1.0 \times 10^{12} \!\!\!\!\!\!\!$
&& $\!\!\!\!\! 426 \!\!\!\!\!\!\!$ & \cr
\omit&height4pt&\omit&&\omit&&\omit&&\omit&\cr
\tablerule
\omit&height4pt&\omit&&\omit&&\omit&&\omit&\cr
&& $\!\!\!\!\! {\rm MW-Ascasibar\ [41] } \!\!\!\!\!\!\!$
&& $\!\!\!\!\! 0.347  \!\!\!\!\!\!\!$
&& $\!\!\!\!\! 1.0 \times 10^{12} \!\!\!\!\!\!\!$
&& $\!\!\!\!\! 421 \!\!\!\!\!\!\!$ & \cr
\omit&height4pt&\omit&&\omit&&\omit&&\omit&\cr
\tablerule
\omit&height4pt&\omit&&\omit&&\omit&&\omit&\cr
&& $\!\!\!\!\! {\rm M31-Klypin\ [39] } \!\!\!\!\!\!\!$
&& $\!\!\!\!\! 0.188  \!\!\!\!\!\!\!$
&& $\!\!\!\!\! 1.6 \times 10^{12} \!\!\!\!\!\!\!$
&& $\!\!\!\!\! 634 \!\!\!\!\!\!\!$ & \cr
\omit&height4pt&\omit&&\omit&&\omit&&\omit&\cr
\tablerule
\omit&height4pt&\omit&&\omit&&\omit&&\omit&\cr
&& $\!\!\!\!\! {\rm M31-Tempel\ [42] } \!\!\!\!\!\!\!$
&& $\!\!\!\!\! 0.661  \!\!\!\!\!\!\!$
&& $\!\!\!\!\! 1.0 \times 10^{12} \!\!\!\!\!\!\!$
&& $\!\!\!\!\! 383 \!\!\!\!\!\!\!$ & \cr
\omit&height4pt&\omit&&\omit&&\omit&&\omit&\cr
\tablerule}}

\caption{Shapiro delays for GRB 070201 from the NFW profiles of the Milky Way and Andromeda using two different data sets.}
\label{Datasets}
\end{table}

The next step is considering a source which is in the Milky Way and estimating the time lag 
by solving Eq.(\ref{cdt}) for $\Delta t$. At this point it is useful to analyze an approximate expression for $\Delta t$
using the previous work of Woodard and Soussa \cite{SW1}. Forcing the dark matter emulators to mimick dark matter isothermal halo distribution, results into the following approximate values for $\Delta A(r)$ and $\Delta B(r)$:
\begin{equation}
A(r)  \approx  1 \; + \;\epsilon \;+  \; \epsilon_{*} \; , \; B(r)  \approx  1 \; - \;\epsilon \;+  \; \epsilon_{*} \ln(\frac{r}{r_s}).
\end{equation}
Here $\epsilon$ is the usual Schwarzschild term $\epsilon \equiv (2GM/c^2r$) and the $\epsilon_{*}$ is the extra small parameter due to this effect, $\epsilon_{*} \equiv 2v_*^2/c^2$ where given the asymptotic rotation speed $v_*$ for Milky Way $\epsilon_{*}$ is at the order of $6\times10^{-7}$. Since $r_s \approx 8 kpc$ the logarithm will be of order one. Looking at Eq.(\ref{cdt}) and observing the direct proportionality of $\Delta t$ with $\Delta A(r)$ and $\Delta B(r)$, one can see that the time delay would approximately be equal to $10^{-6}$ times the total time of flight in flat space. 

\begin{figure}[h!]
\includegraphics[width=3in,height=2in]{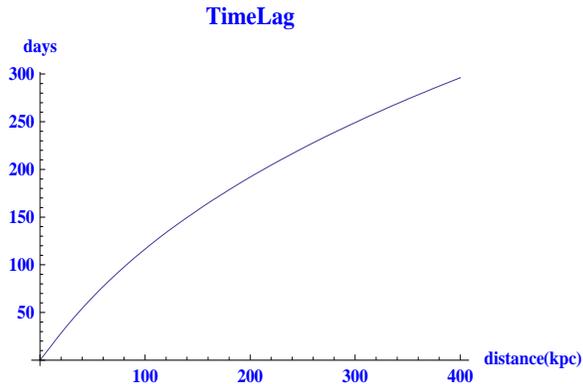}
\caption{Shapiro delays for sources located in Milky Way.}
\label{2d}
\end{figure}

But our aim is to calculate the time lag numerically, due to the difficulty of solving the Eq.(\ref{cdt}) analytically. Using the relevant parameters for the NFW density profile of the Milky Way \cite{KZ},
the numerical solution is depicted in Figure \ref{2d}.

If we look at Figure \ref{2d} we can see that the time lag for an event that is located at 100 kpc is approximately 100 days, which is $10^{-6}$ times $3 \times 10^5$ light years (the total time of flight of a photon in flat space). 

\begin{figure}[h!]
\includegraphics[width=3in,height=2in]{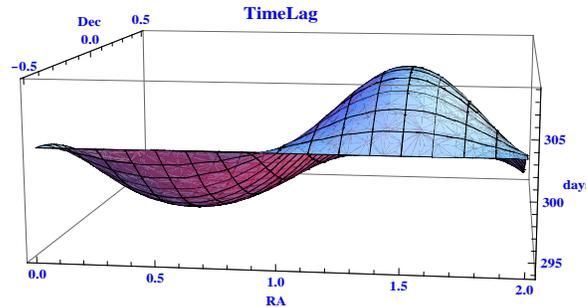}
\caption{The angular dependence of Shapiro delays for sources located in Milky Way. The units of RA(Right ascension) and Dec(Declination) is coverted to radians.}
\label{3d}
\end{figure}

The time lag would surely depend on the angular position of the source, since we are not the center of
the Milky Way. Therefore one would like to see the effect of the direction of the position of a source in the whole sky. 
This was done for a source that is located up to 400 kpc away from the Earth, and it is demonstrated in Figure \ref{3d}.
It can be seen that the angular position of the source might have a maximum of 5\%  effect on the Shapiro delay of the considered source.

\section{Externally Triggered Searches for GWs}

Externally triggered search of gravitational waves is a strategy to look for possible signals around sources which might produce both GWs and other particles(external triggers). The information provided by an external trigger is extremely useful to impose extra requirements for a possible GW signals as well as to increase the confidence of the detection. Knowing the source direction would allow us to look for only relevant portion of the sky. One can even infer the frequency of the possible GW signal based on the information provided by an external trigger and make a data analysis at a specific band \cite{Marka}.

Gamma ray bursts (GRBs) are one of the possible external triggers. A neutron star(NS) black hole or a NS-NS merger might be a good source for a GRB. A highly magnetized NS might also produce short and intense gamma rays, which are called as soft gamma ray Repeaters (SGRs). One sixth of short GRBs are thought to be because of SGRs, where during S5 of LIGO, electromagnetically several hundred of them are were observed \cite{Phi}. Disruption of neutron star's crust would produce an increase in the rotation energy of a neutron star which results in increase of freqency, so called a ``pulsar glitch''.
This might also lead to emission of GWs. Last but not least, supernova explosions would surely produce both GWs and neutrinos. This triggered detection on the other hand, would provide us more information about neutrino mass as well as the mechanism of supernova explosions.  

But still, there has not been a single direct detection of GWs. The search is based on the assumption that, the GW signal and the external trigger are coincident in time within a small time window. Therefore it is possible that, if ``dark matter emulators'' describe gravity then we might be looking at the wrong place. This possibility was suggested before and the possible time lag was calculated for certain events \cite{DKW}. But it is crucial for GW observers (especially the people who are involved with externally triggered search) to know how big this effect is for a particulat object and the amount of error. In this work we provide an estimate of this possible time lag for any object located at Milky Way and current uncertainties are discussed in the next section. 

\begin{table}

\vbox{\tabskip=0pt \offinterlineskip
\def\tablerule{\noalign{\hrule}}
\halign to245pt {\strut#& \vrule#\tabskip=1em plus2em& \hfil#\hfil&
\vrule#& \hfil#\hfil& 
\vrule#& \hfil#\hfil& \vrule#& \hfil#\hfil& \vrule#\tabskip=0pt\cr
\tablerule 
\omit&height4pt&\omit&&\omit&&\omit&&\omit&\cr
&&$\!\!\!\!{\rm M31}\!\!\!\!\!\!\!$ &&$\!\!\!\!\!{\rm M_{tot}(10^{12} M_{\odot})}
\!\!\!\!\!\!$ && $\!\!\!\!\! {\rm Milky\ Way} \!\!\!\!\!$ && $\!\!\!\!\! {\rm M_{tot}(10^{12} M_{\odot})}
\!\!\!\!\!$ & \cr
\omit&height4pt&\omit&&\omit&&\omit&&\omit&\cr
\tablerule
\omit&height4pt&\omit&&\omit&&\omit&&\omit&\cr
&& $\!\!\!\!\! {\rm Corteau\ [46] } \!\!\!\!\!\!\!$
&& $\!\!\!\!\! 1.33^{+0.18}_{-0.18}  \!\!\!\!\!\!\!$
&& $\!\!\!\!\!  {\rm Xue\ [51] } \!\!\!\!\!\!\!$
&& $\!\!\!\!\! 1.0^{+0.3}_{-0.2} \!\!\!\!\!\!\!$ & \cr
\omit&height4pt&\omit&&\omit&&\omit&&\omit&\cr
\tablerule
\omit&height4pt&\omit&&\omit&&\omit&&\omit&\cr
&& $\!\!\!\!\! {\rm Evans\ [47] } \!\!\!\!\!\!\!$
&& $\!\!\!\!\! 1.23^{+1.8}_{-0.6}  \!\!\!\!\!\!\!$
&& $\!\!\!\!\! {\rm Smith\ [52] }  \!\!\!\!\!\!\!$
&& $\!\!\!\!\! 1.42^{+1.14}_{-0.54} \!\!\!\!\!\!\!$ & \cr
\omit&height4pt&\omit&&\omit&&\omit&&\omit&\cr
\tablerule
\omit&height4pt&\omit&&\omit&&\omit&&\omit&\cr
&& $\!\!\!\!\! {\rm Fardal\ [48] } \!\!\!\!\!\!\!$
&& $\!\!\!\!\! 0.74^{+0.12}_{-0.12}  \!\!\!\!\!\!\!$
&& $\!\!\!\!\! {\rm Wilkinson\ [53] } \!\!\!\!\!\!\!$
&& $\!\!\!\!\! 1.9^{+3.6}_{-1.7} \!\!\!\!\!\!\!$ & \cr
\omit&height4pt&\omit&&\omit&&\omit&&\omit&\cr
\tablerule
\omit&height4pt&\omit&&\omit&&\omit&&\omit&\cr
&& $\!\!\!\!\! {\rm Seigar\ [49] } \!\!\!\!\!\!\!$
&& $\!\!\!\!\! 0.73^{+0.02}_{-0.02} \!\!\!\!\!\!\!$
&& $\!\!\!\!\! {\rm Sakamoto\ [54] } \!\!\!\!\!\!\!$
&& $\!\!\!\!\! 1.8^{+0.4}_{-0.7} \!\!\!\!\!\!\!$ & \cr
\omit&height4pt&\omit&&\omit&&\omit&&\omit&\cr
\tablerule
\omit&height4pt&\omit&&\omit&&\omit&&\omit&\cr
&& $\!\!\!\!\! {\rm Ibata\ [50] } \!\!\!\!\!\!\!$
&& $\!\!\!\!\! 0.75^{+0.25}_{-0.13}  \!\!\!\!\!\!\!$
&& $\!\!\!\!\! {\rm Battaglia\ [55] } \!\!\!\!\!\!\!$
&& $\!\!\!\!\! 1.5^{+0.2}_{-0.2}  \!\!\!\!\!\!\!$ & \cr
\omit&height4pt&\omit&&\omit&&\omit&&\omit&\cr
\tablerule}}

\caption{Different mass estimates of the Milky Way and the Andromeda with the corresponding error bars.}
\label{Masses}
\end{table}

\section{Uncertainties in the Calculation}

There are three kinds of uncertainties in the time lag calculation. First one is the position of the assumed source. The uncertainity in the angular position of the GRB's and other sources are determined by a relatively good accuracy. In a previous work For GRB-070201 the error was shown to be of the order of 2\% \cite{DKW}. The second kind of uncertainty is the choice of the dark matter density profile and this choice had an effect which wass less than 1\% \cite{DKW}. The third kind of uncertainty is the data input(parameters) coming from different groups who are working on numerical simulations of dark matter density profiles. And this choice has the biggest effect on the calculation. 

In order to calculate the time lag in Eq.(\ref{cdt}) we need to use these parameters coming from numerical simulations. They enter into the calculation through the mass function. In this work, to get the mass function, we used NFW dark matter density profile for modified gravity models to mimic dark matter. One can see from Table \ref{Datasets}. that using a data set of a different group changed the time lag from 383 days to 634 days for the Andromeda galaxy. And this difference is expected; since the virial mass estimation of M31 is almost twice of one another. 

Therefore the biggest uncertainty comes from the determination of the mass of the galaxies. Looking at Table \ref{Datasets}.  the results for our galaxy are not very different, unlike M31.  But still one naturally ould start wondering about the case of mass estimate for our galaxy as well. Some of the recent mass estimates for both M31 and Milky-Way galaxy are shown in Table \ref{Masses}. The results of virial mass estimates are quite different from one another. This might be misleading since the virial radius which is considered in these works are sometimes different from one another. But even taking that into consideration, the error bars of these estimates are too big, varying from 13\% to 140\% ,that we can not give a precise estimate of the time lag for objects that are far from us. 

To summarize, our prediction of this time lag has certain uncertainties, where the biggest one is the input coming from the numerical simulations of dark matter density profiles of the galaxies. For an event which will occur at a very distant location, more than 100kpc, we can only conclude that the time lag will be big. Therefore the predictivity of this test is much better for an event which is closer to us ($\sim 10kpc$). The errors of the most recent mass estimates are at the order of 20\% and numerical simulation community is hopeful that these mass estimates will get much better in recent future for our galaxy \cite{Ibata}.

\section{Conclusions}

A peculiar property of dark matter emulators is that gravitational waves couple 
to a different metric than the ordinary matter. This results in a time lag between 
the arrival times of these particles coming from the same source, which would be 
a powerful check of these models with a single detection of a gravitational wave. 
A simultaneous detection of gravitational wave and a photon would rule out entire 
class of modified gravity models. But if a time lag is observed which is at the order 
magnitude of the one mentioned in this work, that would mean general relativity is 
wrong and there is no dark matter.

By a numerical analysis we made a rough estimation of the time lag for any source in our galaxy
as far as 400 kpc, and demonstrated it in Figure \ref{2d}. The time lag depends almost linearly on the distance of the source from us, and is approximately $10^{-6}$ times the total time of flight. Estimation of the time lag depends on the data input(mass function of Milky-Way) which is provided by numerical simulations. Therefore the accuracy of this, highly depends on the error bars coming from these simulations. The current state of affairs of the numerical simulations of Milky-Way galaxy puts some constraints on our prediction. Therefore, at this stage we can conclude that the best objects, for this proposed test, are the ones that are close to us($\sim 10$ kpc).

We also have to point that Bekenstein takes the approach as velocity of GWs is smaller than photons \cite{Bek}. But this would lead to an immediate invalidation of TeVeS from the boundary of Moore and Nelson\cite{MN} via gravitational cherenkov-radiation. Therefore we do not agree with Bekenstein's point of view and force the ``dark matter emulators''
to agree with the observed rotation curve and weak lensing results.

Besides its own significance, gravitational wave observation is very important since it will open new windows in our understanding of the universe. One can test whole class of alternate theories of gravity which would results into peculiar imprints such as additional polarization states. Higher order theories of gravity, Brans-Dicke theory, massive graviton theories and Chern-Simons modified gravity are just some of the alternate theories of gravity that we can test (see \cite{Yun,Cap} and references therein). 

This work should serve as a guide for gravitational wave observers who would like to search for imprints of modified gravity models without dark matter. LIGO/VIRGO are currently doing joint science operations at design sensitivities  and searches for gravitational waves with electromagnetic counterparts is a key project \cite{LIGO}. And this test is just another motivation to look for it. 

\vskip 0.1in

\textbf{Acknowledgements}

I am grateful to Richard Woodard and Shantanu Desai for stimulating discussions. 
I would like to thank Szabi Marka for suggesting this problem to me. 
I also would like to thank Hongsheng Zhao and Anatoly Klypin for illuminating comments and discussions.
I am also grateful for the hospitality of the Physics Department of Ko\c{c} University. This
work was partially supported by DFG-Research Training Group "Quantum and
Gravitational Fields" GRK 1523/1, by Marie Curie Grant IRG-247803.

\end{document}